# Enhanced sensitivity of partial O-terminated H-diamond for H$_2$S detection at room temperature

N. Mohasin Sulthana [a,b], K. Ganesan [a,b,1],  P.K. Ajikumar [b]

[a] *Homi Bhabha National Institute, Training School Complex, Anushakti nagar, Mumbai-400094, India*

[b] *Materials Science Group, Indira Gandhi Centre for Atomic Research, Kalpakkam- 603102, India*

## Abstract

The p-type surface conductivity of H-terminated diamond (HD) has opened new path ways to develop diamond based electronic devices, photo-catalysts, chemical and bio-sensors. Herein, we report on the room temperature H$_2$S detection behaviour of pristine HD and the surface modified HD films with partial O-termination (OHD) through ozonation. The response of the pristine HD and partial OHD devices that are ozonated for 30, 60 and 90 s, is found to be ~ 55, 1420, 810 and 95 % respectively, for exposing 900 ppb of H$_2$S under ambient atmosphere at room temperature. Here, the optimally partial OHD sensor displays an enhanced sensitivity by about an order of magnitude due to the catalytic activity of the sparsely populated O-functional groups on HD surface. Moreover, the gas sensor response is found to be higher in wet background atmospheres such as N$_2$ and synthetic air as compared to their respective dry atmospheres. Also, the response curve of these sensors exhibits a peculiar decrease in resistance immediately after exposure to H$_2$S under wet background atmosphere while  such oxidative behaviour is absent under dry atmospheres. Based on these observations, the plausible sensing mechanism of these H-diamond based sensors is proposed with the concept of humidity induced H$_2$S hydrolyzation.

---

[1] Corresponding author. Email : kganesan@igcar.gov.in ( K. Ganesan)

## 1. Introduction

Diamond is one of the promising wide bandgap semiconductors with a lot of attractive properties such as high thermal conductivity, high carrier mobility, high electric breakdown field, chemical inertness, biocompatibility and wide potential window for electrochemical analysis [1,2]. However, the high intrinsic resistivity due to wide bandgap and the lack of shallow dopants are the major issues in conventional device fabrication. Alternatively, H-terminated diamond (H-diamond, HD) behaves as p-type semiconductor with high surface conductivity (SC) which occurs through surface charge transfer doping mechanism [3–5]. This makes HD as an unique candidate for high power and high frequency devices, field effect transistor, electrochemical and photo-catalysts applications [6]. In contrast, the O-terminated diamond (OHD) surface possesses very high intrinsic resistivity [7]. However, the OHD surface exhibits a better catalytic behaviour as compared to HD surface [8]. Consequently, it is intuitive to think that the partial O- termination on HD surface can have the combined advantages of high SC along with O- induced catalytic behaviour. In fact, the FET devices made from partial O-terminated HD surface  demonstrated with added advantages as compared to pristine H-diamond [9]. Hence, the partial O-termination of HD surface will enable to tune the surface electronic properties that helps in performance optimization in the areas like photocatalysis, chemical and bio-sensors. In this article, the advantage of partial O-termination on HD surface is probed for detection of $H_2S$ in ambient conditions at room temperature. We note here that the high SC occurs only within a few top monolayers which contribute to the electrical conduction process of H-diamond during gas sensor measurements while the bulk diamond is electrically inactive due to high intrinsic resistivity ( $> 10^{12}$ $\Omega$cm).

Metal oxide (MOX) based gas sensors are the popular choice for detecting trace level toxic gases because of their fast response, portability, reproducibility and cost effectiveness [10,11]. However, MOX sensors suffer from selectivity among the reducing or oxidizing agents since the charge transfer process is similar for particular type of analyte gases. Also, there are a lot of challenges in achieving room temperature response. In general, MOX sensors work on the principle of charge transfer between the oxide surface and analyte gas through redox reactions. Unlike MOX sensors, HD sensor exhibits a unique water adsorbate layer mediated charge transfer under normal ambient conditions. A gas response happens when reactive gases are adsorbed in the water adlayer and subsequently, they undergo electrolytic dissociation. This electrochemical reactions disrupts the charge equilibrium state of the diamond/water interface and it promotes the changes in the surface conductivity [12]. Such a kind of gas sensing behaviour on H-diamond was first reported for $NO_2$ and $NH_3$ analyte gases with detection limit down to 5 and 2500 ppm, respectively [13–17]. Further, HD sensors exhibit a high degree of selectivity towards a particular group of analyte gas which dissolves in water [18,19]. Also, one more attractive feature of the H-diamond sensor is that it works even at room temperature and also, at humid conditions. Even though $NH_3$ and $H_2S$ both are reducing gases, it is observed that H-diamond is more sensitive to $H_2S$ than that of $NH_3$. Despite a few reports available on the detection of different analyte gases [13–20], the studies on $H_2S$ detection by H-diamond is very limited. Also, the sensor studies based on surface modified H-diamond films are scarce. Furthermore, the sensing mechanism of H-diamond is still not clearly understood. Moreover, unlike a complete coverage of H-atoms on diamond surface, a sparsely distributed O- atoms can exhibit an excellent catalytic behaviour, which can be utilized to enhance the gas sensing response of the H-diamond. This concept motivates to study the effect of O- concentration on the sensing behaviour of H-diamond for $H_2S$.

In this report, the study is focused on to understand the room temperature gas sensing mechanism of H-diamond for $H_2S$ in the ambient condition. Further, we have enhanced the sensitivity of the H-diamond sensors by tuning the surface electronic properties through the substitution of catalytic O- functional groups at H- site. Here, the catalytic O- centres are introduced in an effective and scalable way on H-diamond surface through partial oxidation using ozonation process. In addition, we have also studied to elucidate the humidity effect on this adsorbate mediated $H_2S$ gas sensing of pristine and partially O-terminated HD films.

## 2. Experimental

The microcrystalline diamond films were deposited on $SiO_2/Si$ using ultrahigh pure $CH_4$ and $H_2$ in the ratio of 1: 100 by hot filament chemical vapour deposition, as reported elsewhere [21,22]. Subsequent to growth, the in-situ hydrogen termination was carried out by admitting $H_2$ into the chamber at a flow rate of 200 sccm for 20 min under working pressure of ~ 40 mbar. During H-termination, the substrate and filament temperature were maintained at 800 and ~ 2000 ºC, respectively. After H-termination, the diamond films were cooled down to room temperature in $H_2$ atmosphere. These H-terminated diamond films were exposed to ambient air for 2 days to reach surface charge transfer equilibrium state. Then, to introduce local reactive sites with O-functionalization, these H-diamond films were exposed to ozone atmosphere for 30, 60 and 90 s using UV/ozone pro cleaner (UV ozone cleaner, Ossila). For simplicity, these four samples are labelled as HD, OHD-30s, OHD-60s and OHD-90s, respectively. The effect of H- / O- surface functionalization on the wetting properties of diamond surface was examined by water wetting contact angle (WCA) measurement using telescope-goniometer (APEX, ACAM D1N01, India). Liquid drop analysis (LB-ADSA) was performed using ImageJ software to estimate the WCA. The volume of the water droplet used for WCA measurement is ~ 100 μl. Field emission scanning electron microscope (FESEM, Supra 55, Carl Zeiss, Germany) was used to examine the microstructure and morphology of the as grown diamond film. Raman spectra were recorded by micro-Raman spectrometer (In-via, Renishaw, UK) using diode laser with wavelength of 532 nm and a grating with 1800 grooves/mm. The sheet resistance and sheet carrier density were measured using Hall measurements with Agilent B2902A precision source/measure unit under van der Pauw geometry.

For sensor device fabrication, inter-digitated electrodes with Ohmic contacts were prepared by depositing Pd (20 nm) layer on these H-diamond films by shadow masking method using thermal evaporation. An additional coating of 100 nm Ag layer is provided on the Pd electrode to make thicker electrode with low contact resistance for electrical measurements. The Ohmic contacts were ensured through I-V characteristics. The area, width and inter-electrode distance of the inter-digitated electrode are $\sim 8 \times 8$ mm$^2$, 400 and 300 $\mu$m, respectively. All the gas sensor measurements were performed in static mode at 23 $^0$C and near atmospheric pressure of $\sim$ 960 mbar, unless otherwise mentioned specifically. The schematic of the sensor setup, electrode configuration and the concentration of H$_2$S as a function of time are shown in Fig. 1. Initially, the sensor chamber was filled with background atmosphere such as ambient air; synthetic air (80 % N$_2$ / 20 % O$_2$) and nitrogen gas (100 % N$_2$) under either dry or wet conditions. Subsequently, the analyte gas H$_2$S which was diluted down to 100 ppm in ultra-high pure (UHP) N$_2$ was sent into the sensor chamber through mass flow controller. Systematic sensor measurements were repeated by accumulating H$_2$S inside the chamber. Hence, the concentration of H$_2$S in the chamber was increased cumulatively when the H$_2$S gas was injected into the chamber at periodic interval. The sensor measurements were repeated under different background atmospheres. The relative humidity level was also controlled in the chamber using water bubbler type humidifier. The sensor response curve for four H-diamond based devices (HD, OHD-30s, OHD-60s and OHD-90s) were measured simultaneously using two Agilent B2902A precision source/measure units. The sensor response ( R ) was calculated using the equation,

$$R \ (\%) = (R_g - R_o) * 100 / R_o \qquad \ldots \ldots \ldots \quad (1)$$

where, $R_o$ is the initial resistance and $R_g$ is the maximum resistance of the sensor upon exposure to the analyte gas. The response time of the sensor is defined as the time required to reach 90 % of the resistance value after exposure to the measured gas. Further, the sensitivity of the sensor is defined as the response of the device for a given concentration of analyte gas.

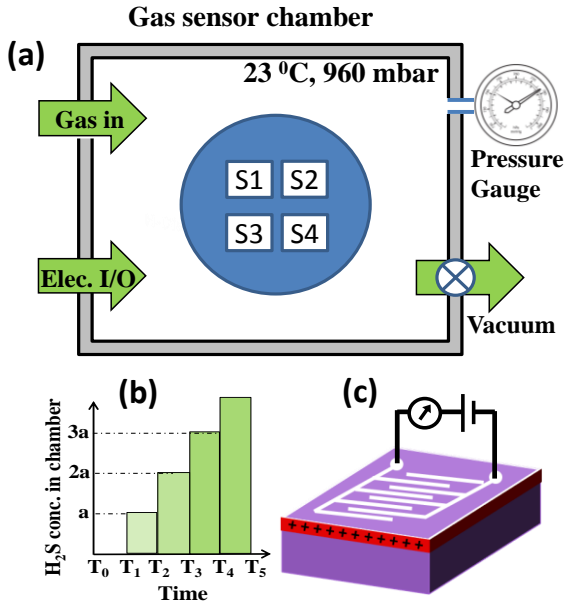

Fig. 1. (a) Schematic of the gas sensing setup for simultaneous sensor measurements for 4 samples, (b) The concentration of $H_2S$ gas in the sensing chamber as a function of time and (c) the inter-digitated electrode configuration used for sensor measurements.

## 3. Results

### 3.1. Structural analysis

Fig. 2a shows the FESEM surface morphology of the as grown H-diamond film. The surface morphology of the diamond film exhibits multi-faceted crystalline structures with grain size of ~ 1.5 µm and the thickness of the diamond film is ~ 2 µm. As can be seen in Fig. 2a, the morphology of the grown diamond film is having triangle and rectangle faceted structures associated with (111) and (100) planes, respectively [22,23]. Fig. 2b shows the Raman spectrum of as grown H-diamond film. It displays a single dominant diamond Raman band at 1332 cm$^{-1}$ with full width and half maximum of ~ 6 cm$^{-1}$. The narrow FWHM of Raman band clearly signify the high structural quality of the polycrystalline diamond film. Since the surface functionalization occurs as a monolayer on diamond surface, the morphology and Raman spectrum do not vary significantly for the pristine and partially ozonated H-diamond films.

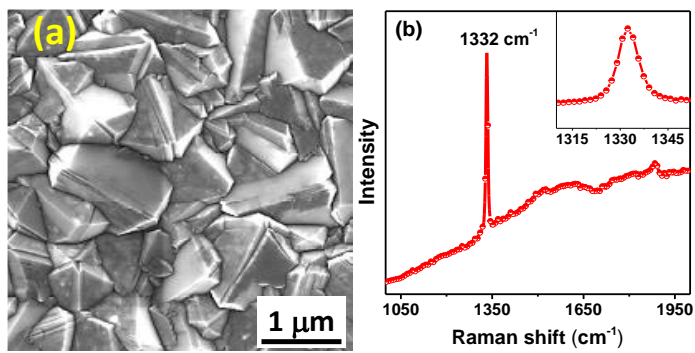

Fig. 2. (a) FESEM micrograph and (b) Raman spectrum of the as grown H-diamond film. The inset in (b) shows the magnified part of Raman band around 1332 cm$^{-1}$.

## 3.2. Wetting contact angle measurements

Fig. 3 displays the variation of water WCA of H-diamond surface after ozonation for specific durations viz. 0, 30, 60 and 90 s. The corresponding optical images of WCA are also given as inset in Fig. 3. The measured WCA of ~ 100° confirms the intrinsic hydrophobic nature of the HD film surface. This hydrophobic nature is due to the dominating van der Waals interaction of the H-terminated surface which weakens the water physisorption [22]. However, this hydrophobic nature transforms into hydrophilic nature, when oxygen functional groups are substituted at H-site on the surface. The polar nature of the oxygen functional groups makes the surface interactive with the polar medium (water). As depicted in Fig. 3, the WCA is found to be 74, 63 and 54° for the samples ozonated for 30, 60 and 90 s, respectively. This gradual transformation of surface wettability affirms the systematic increase of oxygen functional groups on the HD surface.

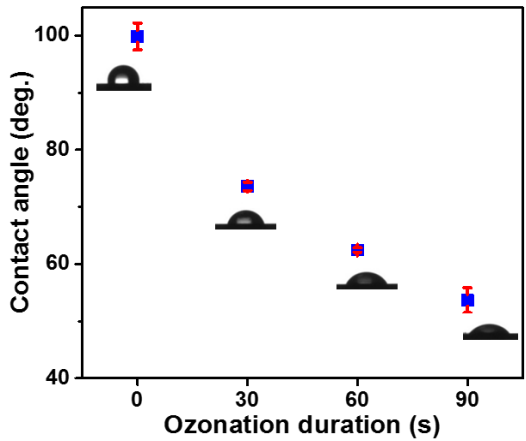

Fig. 3. The variation of water wetting contact angle of H-diamond surface as a function of ozonation duration

### 3.3. Electrical properties

Fig. 4 depicts the I-V characteristics of the HD and partial O-terminated HD films. The linear I-V curves indicates the Ohmic behaviour of the Pd metal electrodes on the films. Fig. 4b shows the variation of sheet resistance as a function of ozonation duration. The sheet resistance of the as prepared HD sample is ~ 8 k$\Omega$/□. This low resistance of HD film confirms the presence of two dimensional (2D) hole accumulation layer on the surface which occurs through surface charge transfer doping mechanism. Further, the sheet resistance of HD films measured after ozonation for 30, 60 and 90 s are found to be ~ 18, 54 & 267 k$\Omega$/□, respectively. Thus, it confirms that HD film sheet resistance increases systematically with substitution of partial O-functional group at H-site. Further, the sheet carrier density of the pristine HD and partial OHD films (30, 60 and 90 s) is found to be ~ 1.1 x $10^{13}$, 4.8 x $10^{12}$, 1.3 x $10^{12}$ and 2.5 x $10^{11}$ $cm^{-2}$ and the mobility of the films is in the range of 60 – 80 $cm^2$/Vs. In partial OHD films, the local electron affinity (EA) changes from negative to positive due to the replacement of H- by O- atoms that minimizes the charge transfer doping. Hence, the surface carrier density decreases about two orders for 90 s ozonation and consequently, the resistance of the partial OHD films increases.

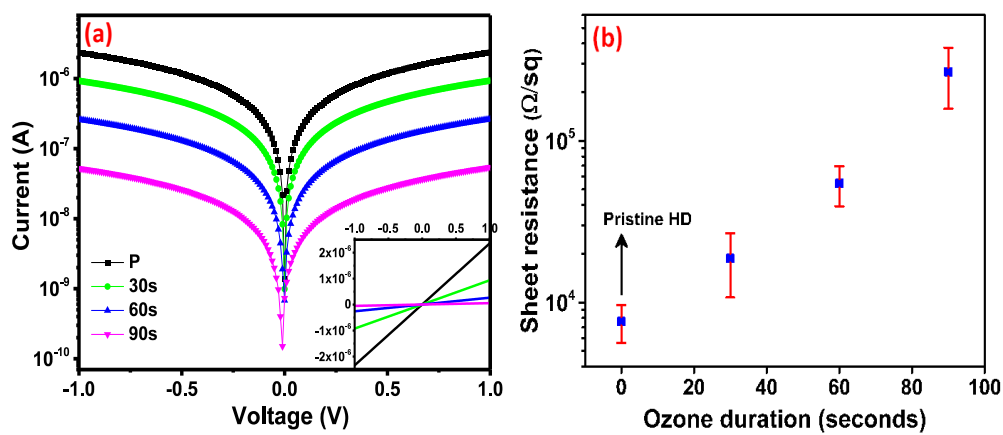

Fig. 4. (a) I-V characteristics in log scale (the inset plot shows in linear scale) and (b) the variation in sheet resistance of pristine and partial O-terminated H-diamond films.

### 3.4. Gas sensing measurements

### 3.4.1. Under ambient air atmosphere (RH ~ 50 %)

Fig. 5a shows the response of pristine and partially O-terminated HD films to the cumulative exposure pulses of $H_2S$ in the concentration range of 0.22 – 0.88 ppm. All the four sensors display an increase in resistance with exposure to $H_2S$, as shown in Fig. 5a. Further, the sheet resistance of the sensors increases as stair case like structure with periodic increase in $H_2S$ concentration in the chamber. The change in resistance resembles the integrator-like gas response for all the four sensors. Even though the baseline resistance of the sensors is different due to the different oxygen coverage on the surface, they do show a similar trend in terms of change in resistance when exposed to $H_2S$. However, the estimated sensor response is strongly dependent on the O-functionalization. All the sensors show a linear response in the concentration range of ~ 0.2 to 0.9 ppm of H$_2$S in ambient air. However, the partially ozonated HD for 30 s (OHD-30s) displays the highest response of ~ 1420 % and the OHD-60s displays a response of ~ 810 % at 0.9 ppm of H$_2$S. The pristine and OHD-90s display the lowest response of ~ 55 and 95 % respectively, for the same H$_2$S concentration.

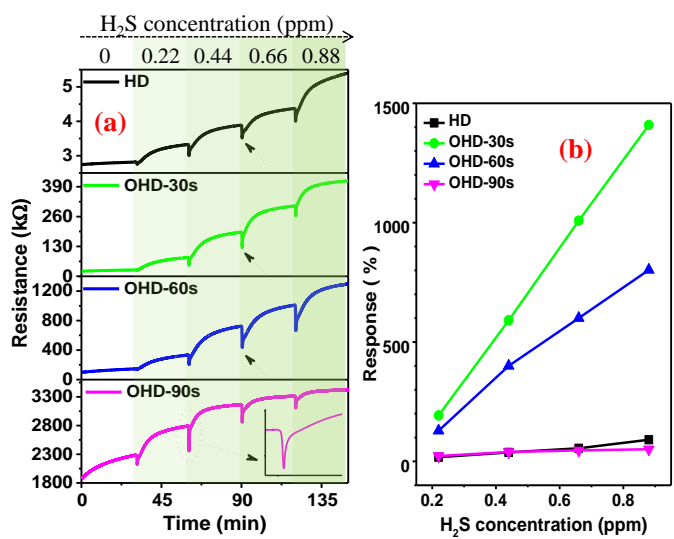

Fig 5. (a) The variation in resistance of pristine and partial O-terminated H-diamond films to the cumulative exposure pulses of H$_2$S. The arrow marks in the curves indicate a sudden decrease in resistance immediately after exposure to H$_2$S gas. (b) The gas sensor response as a function of H$_2$S concentration for these H-diamond based sensors.

Moreover, each sensor shows a sudden decrease in resistance immediately after exposure to H$_2$S, as indicated with arrow marks in Fig. 5a. The reason for sudden change in resistance (peak like behavior) shown Fig. 5a may be thought that it can occur either due to the sudden change in the gas flow on the sensor device or due to some oxidizing reaction taking place on the sensor surface. In order to address this issue, we have also repeated the gas sensing measurements under identical experimental conditions except the background gases viz. synthetic air and N$_2$ under dry and wet conditions and they are discussed below.

### 3.4.2. Under dry conditions in UHP N₂ and synthetic air ( RH < 5% )

The Figs. 6a and 6b show the gas sensor response of the H-diamond based devices to the cumulative exposure pulses of $H_2S$ under dry ( RH < 5 %) $N_2$ and synthetic air atmospheres, respectively. As similar to ambient air atmosphere, the resistance of these sensors increases upon exposure to $H_2S$ in both dry $N_2$ and synthetic air. However, unlike in the ambient air background, the sensors do not exhibit any decrease in resistance upon immediate exposure to $H_2S$. Thus, these sensors show a proper integrator like gas response behaviour towards $H_2S$ exposure pulses under dry $N_2$ and synthetic air atmospheres. As can be seen from Figs. 6c and 6d, the response of these sensors is nearly linear in the concentration range of ~ 0.6 to 2 ppm. However, the highest response is observed for OHD-60s rather than OHD-30s. Also, the sensitivity of the sensors is better in dry synthetic air as compared to dry $N_2$. Furthermore, the sensitivity of the sensors under dry atmosphere is slightly poor as compared to ambient atmosphere with RH ~ 50 %.

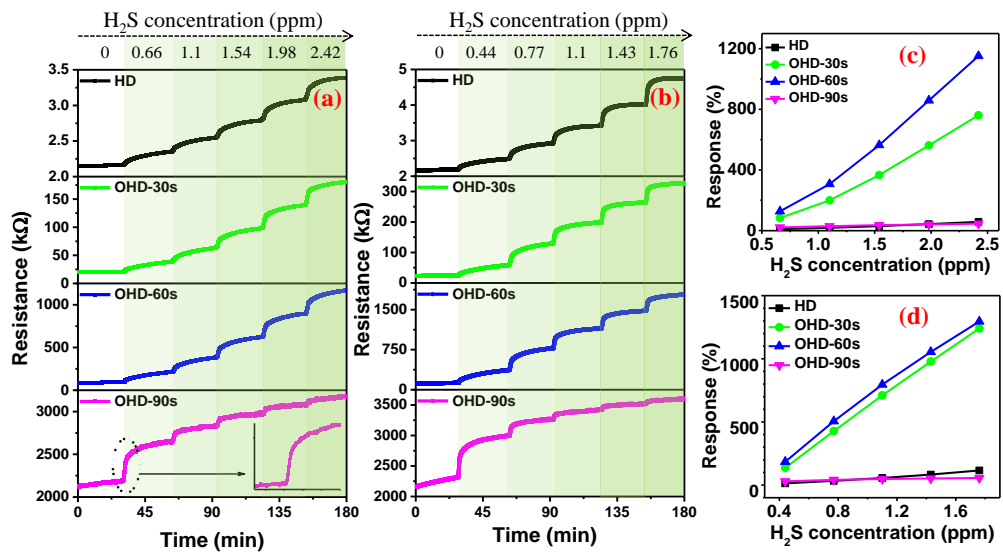

Fig. 6. Sensor response of pristine and partially O-terminated H-diamond films to the cumulative exposure pulses of $H_2S$ gas in (a) dry $N_2$ and (b) dry synthetic air. The gas sensor response as a function of $H_2S$ concentration in (c) dry $N_2$ and (d) dry synthetic air

### 3.4.3 Under humid condition in UHP N₂ and synthetic air ( RH ~ 50 %)

The sensor response of the H-diamond based devices to the cumulative exposure pulses of $H_2S$ under wet $N_2$ and synthetic air medium is shown in Figs. 7a and 7b, respectively. The relative humidity of ~ 50 % was maintained in the sensor chamber through water bubbler humidifier. As similar to ambient air background, the response of these sensors shows a sudden decrease of resistance and subsequent increase in resistance which are due to oxidizing and reducing reactions taking place on the surface, respectively. The presence of oxidizing reaction in wet atmospheres and its absence in dry atmospheres confirm that this oxidizing reaction is due to the interaction of adlayer with the by-product of reactions between $H_2S$ and water molecules in the chamber. Further the sensor response is nearly linear in the concentration range of 0.3 to 1.1 ppm under wet $N_2$ and synthetic air, as shown in Figs. 7c and 7d. As noticed in ambient atmosphere, the OHD-30s shows the highest response of ~ 1120 and 845 % in wet synthetic air and wet $N_2$, respectively. Moreover, these sensor's

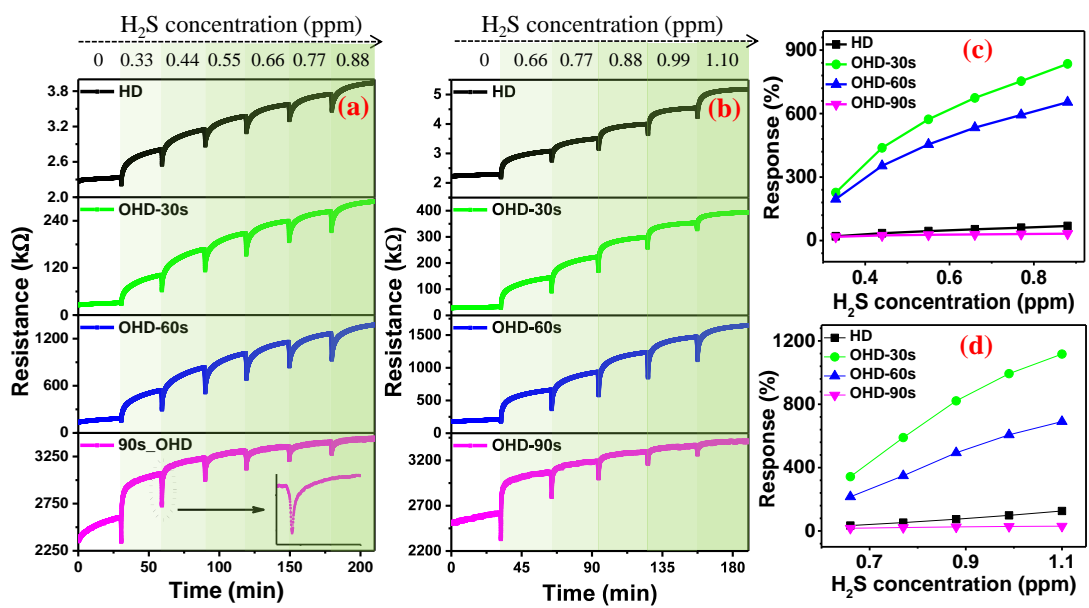

Fig. 7. The change in resistance of H-diamond based sensors to the cumulative exposure pulses of $H_2S$ gas in (a) wet $N_2$ and (b) wet synthetic air. The sensor response as a function of $H_2S$ concentration in (c) wet $N_2$ and (d) wet synthetic air with the relative humidity of ~ 50 %

response is higher in wet atmosphere as compared to that of dry atmosphere. This implies that the humidity plays a significant role on increasing the sensor response.

### 3.4.4. A comparative analysis of sensitivity of H-diamond sensors

Fig. 8 shows the comparison of the sensitivity of the pristine and partial O-terminated HD sensors for ~1 ppm of $H_2S$ in different background atmospheres viz. ambient air; UHP $N_2$ and synthetic air under dry and wet conditions. The bar chart clearly shows that the sensitivity of these H-diamond sensors is lowest under dry atmospheres (both $N_2$ and synthetic air) while it is highest under ambient air atmosphere. Further, the OHD-30s sensor under wet conditions and OHD-60s sample under dry conditions show the highest sensing response among the four samples. The HD and OHD-90s samples show the lowest sensor response among the four in the all atmospheric conditions. This indicates the importance of optimal amount of O- functional groups on the diamond surface. Further, the presence of gaseous oxygen and humidity in the measuring chamber enhance the gas sensitivity of these H-diamond based sensors to $H_2S$.

Fig. 8b shows the response time of the H-diamond sensor for the successive pulses of $H_2S$ analyte gas in the dry and wet synthetic air conditions. It is clearly evident that the response time of the H-diamond sensor decreases gradually from 27 to 12 min in the dry condition for the consecutive pulses of the analyte gas. This indicates that the by-products from the reaction between $H_2S$ gas molecules and adsorbate layer in the dry condition enhances the electron transfer through adsorbate/diamond interface. On the other hand, under wet conditions, such a change in the response time is not observed. These observations indicate that the adsorbate layer thickness on the diamond surface play a significant role on the charge transfer between $H_2S$ and HD surface.

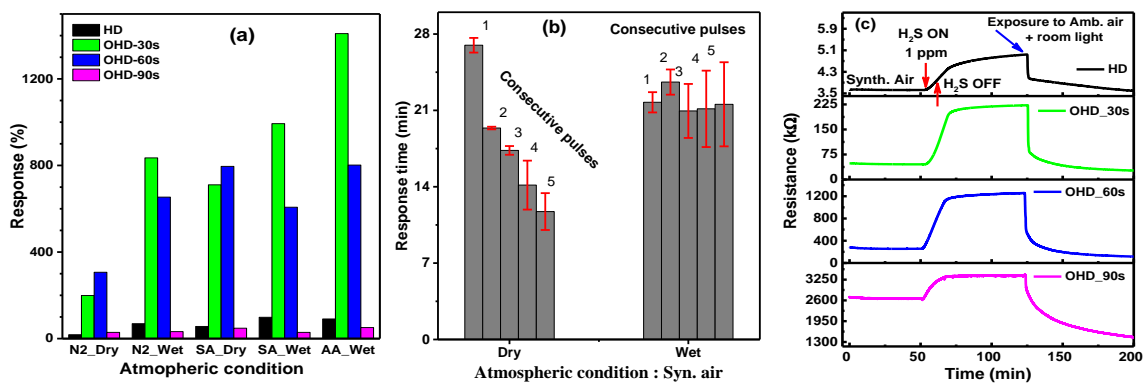

Fig. 8. (a) Comparison of sensitivity of the pristine and partial O-terminated H-diamond sensors for ~ 1 ppm of $H_2S$ measured under different background atmospheres, dry $N_2$ (N2_Dry), humid $N_2$ (N2_Wet), synthetic air (SA) Dry and wet conditions, and ambient air (AA_wet). (b) The response time of the pristine H-diamond sensor to the successive pulses of $H_2S$ gas in dry and wet synthetic air. (c) Response time of the sensors after quick flush of the sensor chamber and then exposure to ambient air and normal room lighting.

Fig. 8c shows the sensor response and recovery of the H-diamond based devices. The sensor response measurement was performed under synthetic air medium under 500 mbar pressure at 23 $^0$C. To study the recovery behaviour of the devices, the sensor chamber was quickly flushed out by ambient air and then, the sensors are exposed to ambient atmosphere with normal room lighting. As can be seen in the Fig. 8c, the sudden decrease in resistance is due to the photocurrent which is induced by the excitation of defects in the diamond. This photocurrent helps to decrease the resistance of the sensors quickly. Subsequently, the $H_2S$ molecules desorbs slowly from the water adsorbate layer of H-diamond surface leading to gradual decrease of resistance of the sensor device to base level. We also note here that the sensor stability and repeatability of the H-diamond based samples are very good even after one month. The base resistance of the devices changes slightly over time. However, the response of the devices does not vary significantly with respect to time delay of a month.

## 4. Discussion

Based on the above mentioned observations, it is noted that these sensors show an increase in resistance for exposing to $H_2S$ in a similar trend under different background atmospheres such as ambient air, both dry and wet $N_2$ and synthetic air. Further, the partial ozonated HD sensors (30 and 60 s) display much higher sensitivity as compared to pristine and ozonated for slightly higher duration of ~ 90 s. Moreover, an unusual decrease in resistance is observed for all the four sensors when the RH was ~ 50 % in the gas chamber irrespective of the background atmosphere. Overall, it is obvious to think that the reducing gas $H_2S$ donates electrons to the p-type HD surface when they interact with each other. Such interaction will lead to increase in resistance of the sensors, as this concept is well established in MOX based sensors. However, a slight change in the surface configuration, i.e. the partial substitution of O at H site, enhances the response about an order which cannot be simply explained by the redox reaction at the surface. In order to understand the sensing behaviour of HD, first we need to understand the origin of SC and how it is affected by partial O-termination and the humidity.

Widely accepted model to explain the SC of HD is surface transfer doping mechanism. When the diamond surface is terminated with H-atoms, the surface C-H bonds create a dipole layer, due to the difference in electronegativity between C and H atoms, that pushes the vacuum energy level to be ~ 1.3 eV below the conduction band minimum [5]. Also, the chemical potential ($\mu_e$) of the adsorbate layer is ~ 4.8 eV which lies below the valence band maximum of the H-diamond. Hence, the spontaneous electron transfer takes place from the diamond valance band to the lowest unoccupied molecular orbital of the adsorbate molecules through electrochemical redox reactions until surface Fermi level coincide with the chemical potential of the adsorbate layer. This electron transfer induces the accumulation of 2D hole gas with upward band bending at diamond surface. Thus, the SC increases enormously due to the induced 2D hole gas at diamond surface [24,25]. As discussed in the electrical measurements, the SC of the diamond is tuned by varying the surface chemistry through ozonation. When the surface H atoms of the diamond are partially replaced by O atoms through ozonation, the chemical and electronic structure change locally due to the formation of C-O related bonds [6]. While the H-diamond surface exhibits NEA, the local EA converts from negative to positive when partial O-atoms substitutes H-atoms. Thus, local EA is tuned by altering the surface coverage of H- and O-atoms. Consequently, the net surface carrier density of these H-diamond films decreases and hence, the resistance increases as shown in Fig. 3. Here, there are two controlling factors that affect the SC of H-diamond viz.  1. EA of the diamond surface and 2. the chemical potential of the adsorbate layer. While the partial O-termination changes the local EA, the analyte gas $H_2S$ and its by-products change the chemical potential and pH of the adlayer and hence, both these factors enhance the gas sensing response of the H-diamond based sensors.

The SC of H-diamond was reported to decrease ( increase ) after exposure to reducing gases such as $NH_3$ and $H_2S$ (oxidizing gases such as $NO_2$ and phosgene) [12,20,26]. Basically, the gas molecules which are capable to dissolve and undergo acid-base reactions in the adsorbate layer could change the SC of diamond. Assuming that the adsorbate layer is neutral and pure, it might contain small and equal densities of $H_3O^+$ and $OH^-$ ions by following water auto-ionization [26].

$2H_2O \longleftrightarrow H_3O^+ + OH^-$ ⋯⋯⋯⋯⋯.. (2)

When the impurity molecules undergo electrolytic dissociation in the adsorbate layer, the local pH and ionic conductivity of the adsorbate layer change depending on the acid-base reaction taking place [8,27]. The electrochemical changes of the adsorbate layer cause the disruption on the already existing charge equilibrium of the diamond/adlayer interface. Suppose the acid forming molecules dissolve in the adsorbate layer, pH of the adsorbate layer decreases that increases the electron transfer from valence band maximum of the diamond to the lowest unoccupied molecular orbital of the adsorbate molecules. It causes more hole accumulation which further increases the SC of diamond surface. Alternatively, if the base forming molecules are dissolved in the adsorbate layer, the pH increases that promotes the electron transfer from adsorbate layer to diamond surface in the opposite direction. Thus, the hole concentration of the diamond surface decreases that eventually decrease SC. Hence any changes in the chemical nature of the adsorbate layer can alter the surface conductivity of diamond.

When H-diamond surface is exposed to $H_2S$ gas molecules, these $H_2S$ molecules are stored into the adsorbate layer in a quasi-permanent manner. The sensing mechanism of HD sensor towards $H_2S$ may be described based on chemical reactions at the air/adlayer and adlayer/diamond interfaces as depicted in Fig. 9. Under dry background atmosphere, a key assumption is that only $H_2S$ molecules are crossing the air/adlayer interface. All adlayer molecules involving in the various chemical reactions are already adsorbed on the HD surface at the beginning and stay adsorbed throughout the process. The following two key reactions are considered in the dissolution reaction of $H_2S$ at the adsorbate layer [28,29].

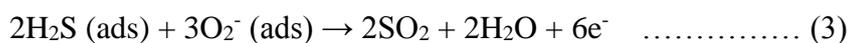

$$2H_2S\ (ads) + 3O_2^-\ (ads) \rightarrow 2SO_2 + 2H_2O + 6e^- \quad \ldots\ldots\ldots\ldots (3)$$

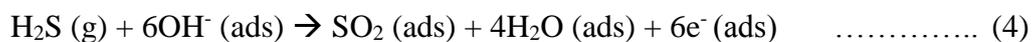

$$H_2S\ (g) + 6OH^-\ (ads) \rightarrow SO_2\ (ads) + 4H_2O\ (ads) + 6e^-\ (ads) \quad \ldots\ldots\ldots.. (4)$$

These reactions lead to change in the pH value and the ionic conductivity of the adsorbate layer. As an outcome of the reactions given in equation 3, the excess electrons flow into the diamond surface and hence, the resistance of the sensors increases.

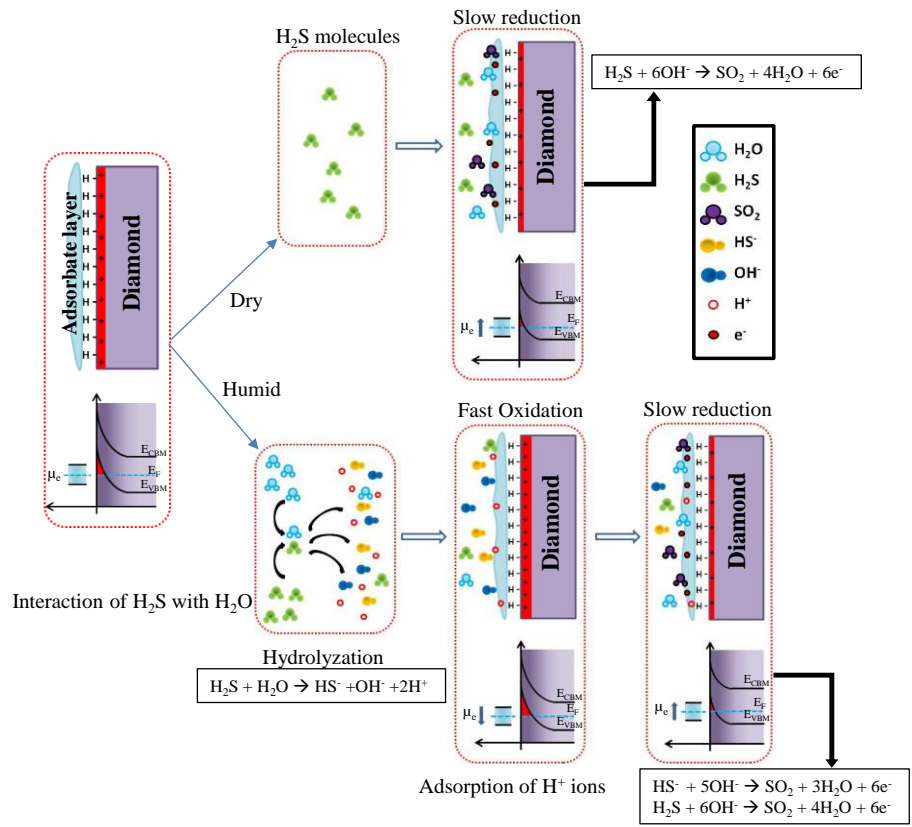

Fig. 9. The $H_2S$ gas sensing mechanism of diamond surface in dry and humid conditions

Under humid background atmosphere, in addition to the regular increase in resistance of these sensor devices, a peculiar oxidation reaction is also observed as shown in inset of Figs. 5a and 7a. This oxidation reaction is caused by the by-product of the interaction between $H_2S$ and $H_2O$ molecules as discussed below. The humidity would induce hydrolyzation of $H_2S$, which would lead to the oxidation reaction in the adlayer/diamond. When it is injected into the chamber with the presence of humidity, the following hydrolyzation reaction would take place [30].

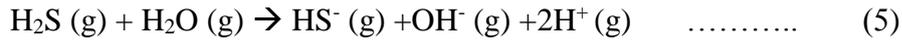

$$H_2S\,(g) + H_2O\,(g) \rightarrow HS^-\,(g) + OH^-\,(g) + 2H^+\,(g) \qquad \ldots\ldots\ldots \qquad (5)$$

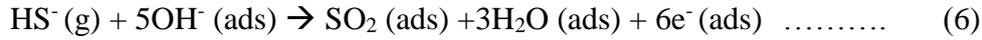

$$HS^-\,(g) + 5OH^-\,(ads) \rightarrow SO_2\,(ads) + 3H_2O\,(ads) + 6e^-\,(ads) \qquad \ldots\ldots \qquad (6)$$

The reaction of $H_2S$ with $H_2O$ molecules would lead to the formation of $HS^-$, $OH^-$ and $H^+$. When these produced $H^+$ ions are getting adsorbed on the adsorbate layer, its pH decreases and ionic conductivity increases and they promote further electron transfer from diamond surface to adsorbate layer. This immediate oxidation process leads to increase in surface conductivity of diamond. In addition, the $H_2S$ molecules which are not interacted with gaseous $H_2O$ molecules would also react with the adsorbate layer in a similar manner like in the case of dry condition as indicated in equation 3. It is noteworthy that the reactions described in both the equations 3 and 6 contribute to decrease the surface conductivity under humid condition. Thus, the sensing response of the H-diamond samples are comparatively higher in the wet condition as compared to the dry condition of the same atmosphere. Thus, the ambient humidity also plays a vital role in the enhancement of sensing properties of the surface conducting diamond films to $H_2S$ at room temperature due to the humidity-induced $H_2S$ hydrolyzation.

One of the important observations in this study is the enhancement of sensor response for the partial O-terminated HD films. In the partial OHD films, a considerable amount of O-functional groups are present, as evidenced by X-ray photoelectron spectroscopy, as reported elsewhere [22]. The local catalytic activity of the surface increases drastically at O-functional groups on the H-diamond surface. When the partial ozonated H-diamond sensor is exposed to analyte gas, the local C-O bonds act as reactive sites to dissolves more amount of gas molecules into the adsorbate layer [31] which alters the pH of the adsorbate layer. Then, these C-O bonds act as active sites to promote electron transfer between adsorbate layer and diamond surface and it changes the conductivity of the neighbouring conductive region [32,33]. Due to these lateral electrostatic interactions, the overall surface conductivity of partial ozonated H-diamond is modulated at higher level than the fully H-terminated diamond surface. Further, it has also been shown that the partial ozonated H-diamond is more sensitive to the change in surface conductivity as a function of pH as compared to fully H-terminated diamond [8]. Hence, the sensor response is high for the partial OHD films such as OHD-30s and OHD-60s. Nevertheless, such an electron transfer is not possible on fully H-terminated diamond surface. That is the reason for a low response for the pristine HD film. At the same time, the presence of a few O-atoms on the HD surface does not affect the NEA of the surface significantly and hence, the surface charge transfer doping is intact. On the other hand, when the O-atoms on the HD surface is beyond a certain critical limit, the local positive EA sites on HD surface increases significantly that affects the global NEA of the surface and hence, the decrease in the surface carrier density. In addition, the surface conducting areas are isolated by the insulating boundaries of oxygen terminated diamond surface. Hence, the sensor response is very poor for the OHD-90s film. Thus, an optimum concentration of O-functional groups on the diamond surface is essential to enhance the response of H-diamond based sensors.

## 5. Conclusions

In summary, the high surface conductivity in microcrystalline diamond films that are grown by hot filament chemical vapour deposition is achieved by in-situ post growth H-termination. The surface conductivity and carrier density of the H-terminated diamond (HD) films were tuned through partial O- surface functionalization. In addition, the surface catalytic activity of the H-terminated diamond (HD) is also increased through O-termination of the pristine HD. Both the pristine and partially O-terminated HD sensors could measure $H_2S$ down to ~ 0.22 ppm in near ambient condition. However, the sensitivity of the optimally partial O-terminated HD is found to be high due to the catalytic behaviour of the additional

O-functional groups on the surface. Moreover, the sensor response is always found to be higher under wet background atmospheres as compared to dry atmospheres. The difference in sensor response under wet and dry background atmospheres is attributed to the humidity induced $H_2S$ hydrolyzation. This study demonstrates the detection mechanism of $H_2S$ down to 0.22 ppm at room temperature using H-diamond based sensors. Also, this study carries the practical importance in developing new chemiresistive sensors based on microcrystalline H-diamond with high sensitivity for $H_2S$ detection at room temperature.

10508. doi:10.1039/c9nj01094g.